\documentstyle[aps,prd,floats,graphics]{revtex}

\def\d{{\mathrm d}} \def\mD{{\cal D}}

\begin{document}
\twocolumn[\hsize\textwidth\columnwidth\hsize\csname@twocolumnfalse\endcsname
\draft

\title{Testing Quintessence models with large-scale structure growth}

\author{K. Benabed, F. Bernardeau}

\address{Service de Physique Th\'{e}orique, CE de Saclay, F--91191 Gif--sur--Yvette
Cedex (France)}

\date{\today{}}

\maketitle
\begin{abstract}
We explore the possibility of putting constraints on quintessence
models with large-scale structure observations. In particular we compute
the linear and second order growth rate of the fluctuations in different
flavors of quintessence scenarios. We show that effective models of
quintessence (e.g. with a constant equation of state) do not account
of the results found in more realistic scenarios. 

The impact of these results on observational quantities such as the
shape of the non-linear power spectrum in weak lensing surveys or
the skewness of the convergence field are investigated. It appears
that the observational signature of quintessence models are specific
and rather large. They clearly cannot be mistaken with a change of
\( \Omega _{0} \). 
\end{abstract}
\textbf{Pacs numbers:} 

] 

\vskip2pc

\section{Introduction}

The recent evidences in favor of a non-zero cosmological constant
\cite{Boomerang:00Nat,Boomerang:ParAnalysis00iprd,Garnavich:SNIa98,Maxima:00,Maxima:ParAnalysis00,Perl:SNIa98ApJ,Perletal:SNIa98Nat,Riess:SNIa98,Waga:00}
have led to developments of alternative scenarios for explaining such
a non-zero vacuum energy density\cite{Weinberg:LambdaPb89}. In particular
the models involving the so-called {}``Quintessence{}'' have attracted
attention from the high-energy physics community\cite{ZlatevSteinhardt:Pipo1998,Tegmark:trucQ01,Steinhardt:Tracker1999,FerreiraJoyce:ExpQuint98,BrxMartin:Quint99,BrxMartin:SUGRA99,RatraPeebles:Quint88}.
Indeed in such models the vacuum energy density is due to the potential
and kinetic energy of a scalar field rolling down its potential. Various
models has been proposed. The simplest implementation of such models,
widely used in the literature, is to introduce an \emph{effective}
quintessence with a constant equation of state\cite{ZlatevSteinhardt:Pipo1998}.
More elaborate theories provide potentials that exhibit a tracking
solution regime as long as the energy density of the quintessence
field is sub-dominant \cite{Steinhardt:Tracker1999}. This behavior
is generically encountered in the Ratra-Peebles\cite{RatraPeebles:Quint88}
model in which the quintessence potential is a simple inverse power
of the field. Other models inspired from high-energy physics have
also been shown to exhibit such a remarkable property\cite{FerreiraJoyce:ExpQuint98,BrxMartin:Quint99,BrxMartin:SUGRA99}. 

The presence of a quintessence field changes the energy content of
the universe and therefore alters its global expansion rate. It is
then natural to try to detect the signature of a non-standard vacuum
equation of state through its impact upon the distance-luminosity
function as it can be revealed by SNIa observations\cite{Starobinsky:QuintRecSNIa99,Astier:2000,Huterer:QuintProb00}.
It has been found however that the precision with which the vacuum
equation of state can be measured depends crucially on whether priors
are assumed on the other cosmological parameters in particular on
the matter content of the Universe. It calls for a re-examination
of the theoretical foundations upon which precision methods for the
determination the cosmic density are based. 

The Cosmic Microwave Background anisotropy power spectrum has been
recognized as a gold mine for the determination of the cosmological
parameters. It is actually a very precious way for measuring the global
curvature of the Universe\cite{Boomerang:00Nat} (through the value
of the angular distance of the last scattering surface) but it suffers
from an unavoidable parameter degeneracy\cite{ZaldaSperSel97,EfstathiouBond99}
so that \( \Omega _{0} \) cannot be determined alone. 

The impact of quintessence models on the properties of the CMB as
well as the primordial density contrast power spectrum has nonetheless
been studied in different models: the \emph{effective} Quintessence
(with the extra shortcoming that the possible fluctuations of the
quintessence field were neglected) as well as some high energy physics
tracking \emph{potential} \cite{FerreiraJoyce:ExpQuint98,RzMartinBrax:CMBQSUGRA2000}.
It has been found that at the redshift of recombination the dark energy
fluid is sub-dominant and has only significant super-horizon fluctuations.
Quintessence effects reveal therefore only as a modest change of the
Sachs-Wolfe plateau, an effect difficult to be unambiguously detected
because of the importance of the cosmic variance. 

It has been shown however that if the intrinsic properties of the
CMB anisotropies fail to provide for an unambiguous test of quintessence,
its existence can be betrayed by the amplitude of the density fluctuations
on the last scattering surface compared to those at low redshift.
This can be done for instance with the help of galaxy cluster counts\cite{Ekeetal96,CMBsigma8forQ}
or with weak lensing measurements\cite{Maoli:lensSurv00,vWetal:lensSurv01}. 

In all cases however direct constraints of \( \Omega _{0} \) that
would help to disentangle models rely on analysis of the local universe
properties. Matter content of galaxy clusters or their number density
evolution \cite{Ekeetal96,OukbirBlanchard97} can provide useful constraints.
Unfortunately these methods depend on non-trivial modeling of cluster
properties such as X-ray luminosity or temperature--mass relations.
It is therefore unlikely they can provide accurate constraints on
\( \Omega _{0} \) with a well controlled level of systematics. 

New methods, based on weak lensing observations, are now emerging
that are in principle free of elaborate physical modeling. The proposed
means for constraining \( \Omega _{0} \) are based on the rate at
which non-linear effects start to play a role in the cosmic density
field. Fundamentally two ideas have been followed. One is based on
the nonlinear evolution of the shape of the power spectrum\cite{JainSel97}
and preliminary results have already been reported in this case\cite{vWetal:lensSurv01}.
It relies here on some specific class of models namely some flavor
of CDM model (the shape of the nonlinear power spectrum depends obviously
on what is assumed for the linear one). The other method, proposed
in \cite{BvWM:WLensTheo97}, is more demanding on the observation
side but is based on the only assumption that the initial conditions
were Gaussian. It relies on the direct detections of non-Gaussian
properties of the density field. In particular it has been shown that
the large-scale convergence skewness can be used to measure \( \Omega _{0} \).
Exact results obtained via second order perturbation theory have been
obtained for models with or without a cosmological constant. It is
then crucial to know whether these results would be affected in case
of quintessential dark energy. 

The aim of this paper is therefore to examine the growth of structure
in both the linear and the nonlinear regime. To illustrate our results
and their robustness we consider various realistic models of Quintessence.
So far the evolution of large scale structure has only been studied
with \emph{effective} quintessence \cite{ZlatevSteinhardt:Pipo1998,Hui:S3lech99,Ma:QuintSpectrePuiss99}
and we will see that it does not provide a realistic account of what
is happening in explicit models of quintessence. 

This paper is divided as follows, in Section II we describe the models
we use and in particular the evolution of the vacuum equation of state
they imply. In Section III the results for the linear and second order
growth rate are presented. Implications of these results are discussed
in section IV for the nonlinear power spectrum.

\section{The Quintessence models}

We postulate that the content of the universe includes a scalar field,
\( Q \), of potential \( V \). This scalar field is responsible
for the dark energy we observe today and usually described as a cosmological
constant. Its motion equation is given by the Klein-Gordon equation
\begin{equation}
\label{KG}
\ddot{Q}+3H\, \dot{Q}=-\frac{\partial V}{\partial Q}
\end{equation}
 and it contributes to the energy and pressure terms with\begin{eqnarray}
\rho _{Q} & = & V(Q)+\frac{1}{2}\dot{Q}^{2},\nonumber \\
p_{Q} & = & -V(Q)+\frac{1}{2}\dot{Q}^{2}.
\end{eqnarray}
 The equation of state of the dark energy \begin{equation}
p_{Q}=\omega _{Q}\, \rho _{Q}
\end{equation}
 is \emph{a priori} no longer characterized by a constant \( \omega _{Q}=-1 \)
parameter. It can vary from \( \omega _{Q}=-1 \) when the dynamic
of the field is dominated by its potential, to \( \omega _{Q}=1 \),
when the kinetic energy dominates. In all the models we will consider,
the parameters will be chosen so that \( \Omega _{0}=0.3 \) and \( \Omega _{Q}=0.7 \)
today, except otherwise mentioned.

In the following we focus our analysis on two models with tracking
solutions that provide explicit time dependency of the equation of
state, the Ratra-Peebles model\cite{RatraPeebles:Quint88} and the
model developed by Brax and Martin in which the potential shape incorporates
generic Super--Gravity factors\cite{BrxMartin:Quint99,BrxMartin:SUGRA99}.
In this section we succinctly review the properties of the \emph{effective},
Ratra-Peebles and SUGRA quintessence models and compute the resulting
equation of state of the Universe in these models.

\subsection{\emph{Effective} Quintessence}

Models of \emph{{}``effective{}''} quintessence are the simplest
implementations of a non-trivial vacuum equation of state. It is simply
assumed that the equation of state parameter is fixed and represent
an average value of a cosmic component following a complex evolution
(whether it is a quintessence field or not). Considered as a simplified
version of a quintessence model, this is a valid approach if, for
some reasons, the kinetic energy and the potential are almost constant
and of the same order. This condition (which is \emph{not} the slow-roll
condition where the kinetic energy is much smaller than the potential)
seems unlikely to be verified in a realistic framework. We will nonetheless
compare realistic models with this approximation to show its impact
on observed quantities.

\subsection{Tracking Quintessence}

In a very wide class of quintessence models the field dynamics exhibits
a tracking solution. It is such that the evolution of the dark energy,
during radiation and matter domination, is completely determined by
the potential shape regardless of the initial conditions\footnote{%
over hundred orders of magnitude 
}. In other word, the only tuning they require to reproduce today's
observations is the energy scale of the potential. Eventually, this
scale will have to be explained by high energy physics computations.
While this task seems insuperable in case of a pure cosmological constant,
it might be within theoretical grasp for tracking quintessence\cite{Steinhardt:Tracker1999,Binetruy:1998}

\begin{figure}[ht]
{\centering \resizebox*{8cm}{!}{\includegraphics{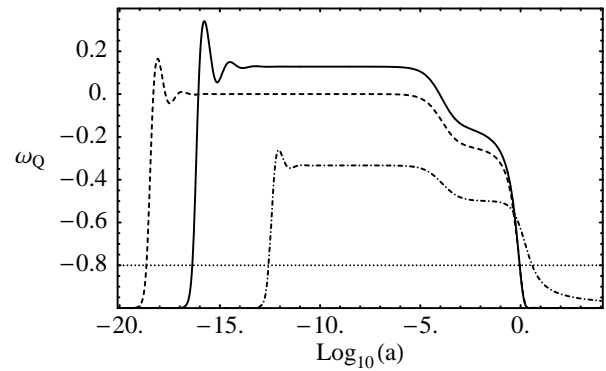}} \par}

\caption{The evolution of the vacuum equation of state as a function of the
expansion parameter \protect\( a\protect \) for different cosmological
models. The dotted line corresponds to a vacuum equation of state,
\protect\( p=-0.8\rho \protect \); the dot-dashed line to a Ratra-Peebles
solution with \protect\( \alpha =2\protect \); the dashed line to
a SUGRA behavior with \protect\( \alpha =6\protect \) and the solid
line to \protect\( \alpha =11\protect \). The amplitude of the quintessence
potentials is such that \protect\( \Omega _{\textrm{matter}}=0.3\protect \)
at \protect\( z=0\protect \) in all cases. }

\label{omall}
\end{figure}

The phenomenological properties of such class of models can be summarized
through the time evolution of the cosmic equation of state. The detailed
behavior of the field in the first stages of its evolution depends
on the initial conditions. If initially \( \rho _{Q} \) represents
a fair fraction of the cosmic energy density, the field \( Q \) rolls
quickly down its potential so that the quintessence energy density
is purely kinetic. It is slowed by the expansion until it freezes
at a value larger than the one corresponding to the tracking solution.
The value of the field remains then constant ---the quintessence energy
density is purely potential--- until it catches with the attractor
solution. Once on the attractor solution, the equation of state parameter
of the quintessence field takes a value that depends only on the shape
of the potential and on the equation of state of the dominant specie
of the universe (it therefore changes at equivalence). When the energy
density of the field starts to dominate, the field follows an inflationary
type slow roll solution whose equation of state is closing to \( \omega _{Q}=-1 \).
These behaviors are displayed on Fig. \ref{omall} for the potentials
we adopted. The time at which the tracking solution is reached is
completely arbitrary and has no effects on the quantities we consider
in the following. 

Our analysis will be done for two tracking models: 

\begin{itemize}
\item the Ratra-Peebles model \cite{RatraPeebles:Quint88} whose potential
is\begin{equation}
\label{RP}
V_{\textrm{RP}}(Q)=\frac{M^{4+\alpha }}{Q^{\alpha }}
\end{equation}
 and which is the simplest model exhibiting a tracking solution. In
particular, it is very hard, with this potential, to get an equation
of state \( \omega _{Q}<-0.7 \) while keeping a reasonable (from
the high energy physics point of view) energy normalization for \( M \)
if \( \Omega _{\Lambda }=0.7 \) today. Note that for such a potential
the vacuum equation of state of the attractor solution is given by
\begin{equation}
\label{omegaQtrack}
\omega _{Q}=\frac{-2+\alpha \omega _{B}}{\alpha +2}
\end{equation}
 where \( \omega _{B} \) is the equation of state parameter of the
background fluid (\( 1/3 \) for a radiation dominated universe, 0
for a matter dominated universe). In the following we will consider
the case \( \alpha =2 \), which gives \( \omega _{Q}\sim -0.6 \)
today, marginally consistent with the supernovae observations although
it leads to an unrealistic low energy scale for \( M \). 
\item the SUGRA model, proposed by Ph. Brax and J. Martin \cite{BrxMartin:SUGRA99,BrxMartin:Quint99}
whose potential is\begin{equation}
V_{\textrm{Sugra}}(Q)=\frac{M^{4+\alpha }}{Q^{\alpha }}\, \exp \left[ 4\pi \frac{Q^{2}}{M_{\textrm{Planck}}^{2}}\right] .
\end{equation}
 The corrective factor is motivated by the fact that, in the Ratra-Peebles
scenario, the field naturally reaches the Planck scale at low redshift.
If the quintessence potential is to be derived from models beyond
the standard model of particle physics that are expected to include
super-gravity properties, it is natural to expect super-gravity corrections
in the shape of the potential. The potential Ph. Brax and J. Martin
proposed is actually the extension of the Ratra--Peebles potential,
with a generic Super--Gravity correction (the exponential term). This
last model is of particular interest since its predictions are in
good agreement, for a wide range of parameters, with the SNIa measurements.
We studied here two examples of this potential, \( \alpha =6 \) and
\( \alpha =11 \) which both lead to equation of state \( \omega _{Q}\sim -0.8 \)
at zero redshift. These choices of parameter lead to an energy scale
\( M \) from \( 10^{6} \) to \( 10^{11} \) GeV, which does not
contradict our knowledge of high energy physics. 
\end{itemize}
The two models have the same tracking solution and the equation of
state parameter is thus the same, given by Eq. (\ref{omegaQtrack}),
on it. Differences between two models arise when the field leaves
the tracking solution. At this time, the field value is of the order
of the Planck mass, and the SUGRA correction of the latter models
starts to dominate. This SUGRA correction cures the problems encountered
by the Ratra-Peebles potential by quickly slowing the field as it
rolls down thus providing a smaller equation of state parameter\cite{BrxMartin:SUGRA99,BrxMartin:Quint99}. 

\subsection{Solution of the equations with Quintessence}

The quintessence field contributes to the Friedman equations, and
therefore to the evolution of the expansion rate of the universe,
\begin{eqnarray}
\left( \frac{\dot{a}}{a}\right) ^{2} & = & \frac{8\pi }{3M_{\textrm{Planck}}}\rho _{\textrm{tot}.}\\
\frac{\ddot{a}}{a} & = & -\frac{4\pi }{3M_{\textrm{Planck}}}\left( \rho _{\textrm{tot}.}+3p_{\textrm{tot}.}\right) 
\end{eqnarray}
 where \( \rho _{\textrm{tot}.} \) is the total energy density of
the Universe and \( p_{\textrm{tot}.} \) its pressure assuming we
live in a zero curvature universe. It is convenient to define the
parameter \( \omega  \) as the effective equation of state parameter
of the ensemble of the cosmic fluids, \begin{equation}
p_{\textrm{tot}.}=\omega \, \rho _{\textrm{tot}.}.
\end{equation}

This parameter is expected to vary from \( 1/3 \) in the radiation
dominated era, \( \omega =0 \) in the matter dominated era to \( \omega \rightarrow -1 \)
when the vacuum energy dominates. The shape of this transition, and
its implication on the growth of structure is precisely what we investigate
in this paper.

\begin{figure}[ht]
{\centering \resizebox*{8cm}{!}{\includegraphics{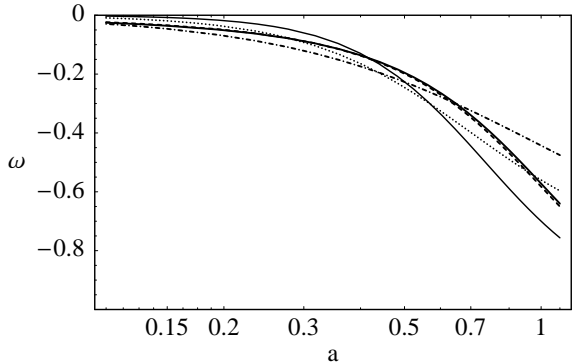}} \par}

\caption{The evolution of the global cosmic equation of state as a function
of the expansion parameter \protect\( a\protect \) for different
cosmological models for a redshift range, \protect\( z=0\protect \)
to \protect\( 10\protect \). Same convention as Fig. \ref{omall}
with thin solid line corresponding to a pure cosmological constant
such that \protect\( \Omega _{\Lambda }=0.7\protect \) at \protect\( z=0\protect \).}

\label{omegaall}
\end{figure}

The evolutions of the equation of state of the universe are shown
on Figs \ref{omall}-\ref{omegaall}. They show that the transition
from a matter dominated universe \( \omega =0 \) to a vacuum dominated
universe is much smoother in case of Quintessence models. In fact,
the universe leaves the \( \omega =0 \) line much sooner in the tracking
quintessence models than in the \emph{effective} quintessence, or
the \( \Lambda  \) models. It is then natural to expect significant
effects on the angular distances or on the growth of fluctuations.
Moreover, the low redshift behavior of the global cosmic equation
of state is very different in the three quintessence models. This
also should induce significant phenomenological differences between
the \( \Lambda  \) models and the quintessence models.

\begin{figure}[ht]
{\centering \resizebox*{8cm}{!}{\includegraphics{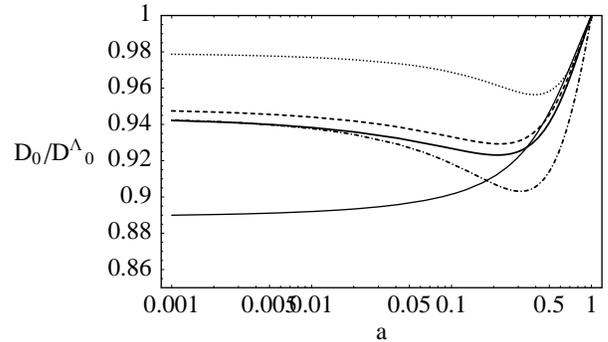}} \par}

\caption{Evolution of the comoving angular distance in the different scenarios
compared to a pure cosmological constant from \protect\( z=1000\protect \)
to the local universe. Same conventions as in Figs. 1-2 are used for
the line styles. The lower solid line corresponds to a pure cosmological
constant model with \protect\( \Omega _{0}=0.4\protect \).}

\label{distances}
\end{figure}

The implication of these behaviors for the angular distances is shown
on Fig. \ref{distances}. The differences seem not very noticeable
at small redshift. However, they build up to be significant when the
effect is integrated to the last scattering surface. The end values
of the angular distances for the different tracking quintessence scenarios
are very close to each other, although they correspond to different
values of the equation of state today (see fig \ref{omegaall}). Not
surprisingly quantities sensitive to the angular distance at high
\( z \), such as the position of the first acoustic peak, have been
found to depend upon the vacuum equation of state\cite{RzMartinBrax:CMBQSUGRA2000}.
One should also expect significant differences in the amplitude of
the lens effect on the CMB anisotropies (as its depends on the angular
distances between the observed objects and the lenses). We note however
that a \( \Lambda  \) model with a lower value of \( \Omega _{0} \)
can reproduce fairly well the low redshift behaviour of the angular
distances in our Quintessence models.

\section{The growth of structure}

\subsection{The linear growth rate}

\label{sec:lingrowth}

In the previous section we have observed that the evolution of the
universe with a quintessence component being smoother than with a
pure cosmological constant, its departure from the case of an Einstein-de
Sitter (EdS) universe occurs later in the former. This effect has
been described before \cite{Ma:QuintSpectrePuiss99,ZlatevSteinhardt:Pipo1998,Tegmark:trucQ01}
however only in the context of \emph{effective} quintessence, but
it is clearly amplified here because the energy fraction of the quintessence
field can be much larger at high redshift in cases of realistic quintessence
models. 

In this section we investigate the impact of these effects upon the
evolution of large scale structure. We only consider the large scale
structure history after recombination, a time at which the dark matter
fluctuations dominate. After recombination and at sub-horizon scales
the quintessence field perturbations correspond to decaying modes
and can therefore be ignored. Within these assumptions the growth
rate of the density contrast at linear order is driven by the equation
\cite{Peebles:LargeScale80}, \begin{equation}
\ddot{D}_{1}(t)+2H\dot{D}_{1}(t)-\frac{3}{2}H^{2}\, \Omega (t)\, D_{1}(t)=0
\end{equation}
 where \( \Omega (t) \) corresponds to the fraction of energy in
the matter component. The growth rate is independent on the wavelength
of the fluctuations as long as we consider fluctuation at sub-horizon
scale and if we neglect the pressure effects. The time evolution of
\( D_{1} \) provides the amplitude of the density fluctuations.
\begin{figure}[ht]
{\centering \resizebox*{8cm}{!}{\includegraphics{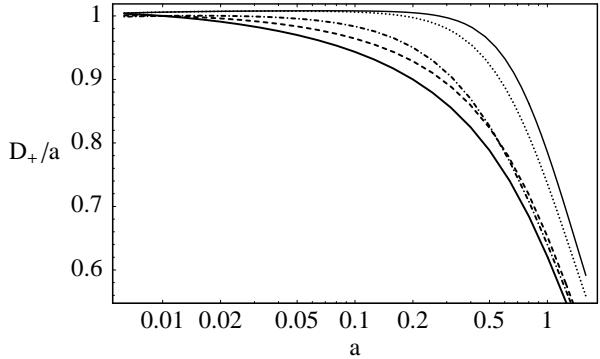}} \par}

\caption{The ratio \protect\( D_{+}/a\protect \) (normalized to unity at
\protect\( z>100\protect \)) for the different scenarios. The today
growth rate is smaller by about 20\% in tracking quintessence scenarios.}

\label{dplus}
\end{figure}
 Fig. \ref{dplus} gives the growth rate, for different models, as
a function of redshift compared to the growth rate in EdS\footnote{%
whose solution is well known \( D_{1}(t)=a(t) \) 
}. In models with a pure cosmological constant, the growth factor remains
very close to the EdS solution for a long period, then changes abruptly
between redshift 2 and 3. In quintessence models the solutions follow
the same scheme, yet with a smoother slope change. However, for the
same normalization at \( z>100 \), when the Universe is very close
to a Einstein-de Sitter model, the today's growth rates are quite
different. The quintessence scenarios with a tracking field exhibit
a smaller growth today (of order 20 to 30 percent less). To say it
in other words, for the same \( \sigma _{8} \), the quintessence
models demand for larger density fluctuations at early times. This
effect should lead to a difference between CMB normalizations and
low redshift normalizations. 

The origin of this difference is clear. It is due to the fact that
the energy fraction in the quintessence field remains significant
for a much longer time. With this respect the \emph{effective} quintessence
solution reveals very similar to the \( \Lambda  \) scenario, whereas
realistic models of quintessence leads to linear growth rates that
depart from the EdS case at redshift as large as 30! Clearly, models
of \emph{effective} quintessence that can match the SNIa observations
do not provide a good account of the linear growth rates found in
realistic models of quintessence.

\subsection{Second order growth rate}

As mentioned in the introduction, for Gaussian initial conditions
the second order growth rate determines the rate at which non-Gaussian
properties emerge in the matter density field. In a perturbation theory
approach it is indeed this quantity that determines the value of the
large-scale skewness. Furthermore, it turns out that weak lensing
surveys can be used as a test-ground for this effect and would provide
a robust constraint on \( \Omega _{0} \) through the value of convergence
skewness\cite{vWBM:WLensSim99,BvWM:WLensTheo97}. We examine here
to what extent this approach remains valid in a quintessence cosmology. 

Such perturbation theory calculations are based on the computation
of higher order terms in a perturbative approach. More precisely the
reduced skewness defined as, \begin{equation}
S_{3}=\frac{\langle \delta ^{3}\rangle }{\langle \delta ^{2}\rangle ^{2}}
\end{equation}
 can be related to the second order growth rate\cite{BouchetJuszkColombi:skewCollapse92}
and more specifically to the second order growth rate in the spherical
collapse dynamics\cite{Bernardeau92,Bernardeau:SkewKurto94}. In this
case it is simple to expand the local density contrast to second order
with respect to the initial density fluctuations, \begin{equation}
\delta _{\textrm{sc}}(t)=D_{1}(t)\, \delta _{i}+\frac{D_{2}(t)}{2}\, \delta _{i}^{2}+\ldots 
\end{equation}
 with time dependent coefficient that can be explicitly calculated
for any cosmological model. The function \( D_{2}(t) \) is the growing
mode of the equation, \begin{eqnarray}
\ddot{D}_{2}(t)+2H\dot{D}_{2}(t)-\frac{3}{2}H^{2}\, \Omega (t)\, D_{2}(t)= &  & \nonumber \\
 &  & \hspace {-3cm}3H^{2}\, \Omega (t)\, D_{1}^{2}(t)+\frac{8}{3}\dot{D}_{1}^{2}(t).
\end{eqnarray}
 The 3D density skewness at large scale (and when smoothing effects
are neglected) is then directly proportional to \( D_{2}(t) \), and
is given by \begin{equation}
\label{skewness3D}
S_{3}=3\, \frac{D_{2}(t)}{D_{1}^{2}(t)}.
\end{equation}

\begin{figure}[ht]
{\centering \resizebox*{8cm}{!}{\includegraphics{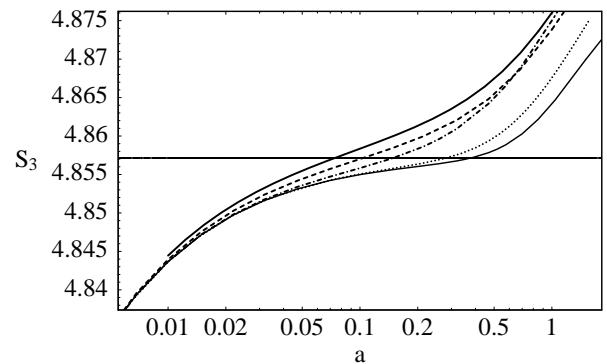}} \par}

\caption{The skewness for the different scenarios. The horizontal straight
line gives the skewness in EdS models. The variations on the quintessence
models are very small. }

\label{skewness}
\end{figure}

From Fig. \ref{skewness} it is clear that the variations of the skewness
with the cosmological models are very small (below percent level)
and are likely to remain undetectable. It means that the second order
growth rate does not introduce further dependence with the vacuum
equation of state. This result actually extends a property already
known for the dependence of \( S_{3} \) on \( \Omega _{0} \) for
open universes\cite{BouchetJuszkColombi:skewCollapse92}, flat universes
with a cosmological constant\cite{Bernardeau:SkewKurto94} or in some
flavors of non standard vacuum equation of state\cite{GaztaLobo01}. 

The skewness of the convergence is thus expected to be left unchanged,
except through the dependence of the angular distances and the linear
growth rate of the fluctuations. We recall here the formal expression
of the convergence skewness in perturbation theory for a power law
spectrum (of index \( n \))\cite{BvWM:WLensTheo97}. For a flat Universe,
it is given by, \begin{equation}
\label{s3kappa}
s_{3}=\frac{\int _{0}^{\mD _{s}}\d {\mD }\, w^{3}(\mD )\, D_{1}^{4}(\mD )\, \mD ^{-2(n+2)}\left[ S_{3}^{2D}-\frac{3}{2}(n+2)\right] }{\left[ \int _{0}^{\mD _{s}}\d {\mD }\, w^{2}(\mD )\, D_{1}^{2}(\mD )\, \mD ^{-(n+2)}\right] ^{2}}
\end{equation}
 with an efficiency function \( w(\mD ) \) given by \begin{equation}
\label{eq:w:def}
w(\mD )=\frac{3}{2}\, \Omega _{0}\, \frac{\mD (\mD _{s}-\mD )}{\mD _{s}\, a}
\end{equation}
 where \( \mD _{s} \) is the comoving distance to the sources and
\( s_{3} \) is the skewness parameter for the 2D dynamics. The latter
can be related to the 3D one, since it implies only a different combination
of the terms appearing for the 3D case\cite{Bernardeau:SkewKurto94,BvWM:WLensTheo97}.
It finally gives, \begin{equation}
s_{3}=\frac{3}{2}+\frac{9}{4}\frac{D_{2}}{D_{1}^{2}},
\end{equation}
 which amounts to \( 36/7 \) for an Einstein-de Sitter case.

\begin{table}[ht]

\caption{\label{tables3conv}Value of the skewness, Eq. (\ref{s3kappa}) of
the local convergence in weak lensing surveys for sources at redshifts
1 or 2 and for a power law index \protect\( n=-1.5\protect \).}

\vspace{.1cm}

\begin{tabular}{ccc}
 skewness &
 \( z_{s}=1 \)&
 \( z_{s}=2 \)\\
\hline
\( \Lambda  \) model (\( \Omega _{0}=0.3 \)) &
 76. &
 26. \\
 \( \Lambda  \) model (\( \Omega _{0}=0.25 \)) &
 85. &
 28. \\
 \( w_{Q}=-0.8 \)&
 83.&
 28. \\
 Ratra-Peebles, \( \alpha =2 \)&
 91. &
 32. \\
 Sugra, \( \alpha =6 \)&
 85. &
 30. \\
 Sugra, \( \alpha =11 \)&
 86. &
 30. \\
\end{tabular}
\end{table}

In table \ref{tables3conv} we present the expected skewness for the
different models we have considered for sources at redshifts 1 or
2. The results show that the projection effects on the value of the
skewness can be pretty large. They increase the value of the skewness
so that Quintessence models with \( \Omega _{0}=0.3 \) mimic what
one expects for a \( \Omega _{0}=0.25 \) model with a pure cosmological
constant. 

It is to be noted that angular diameter distances, in a quintessence
scenario, rather resemble a \( \Lambda  \) model with a \emph{larger}
value of \( \Omega _{0} \) (see Fig. \ref{distances}). From those
two joint observations it should then be possible to test the quintessence
model hypothesis. Results should however be extended to the intermediate
and nonlinear regime where most of the data are going to be although
we expect the qualitative results found here to remain valid.

\section{Non-Linear Matter Power Spectrum in Quintessence models}

\subsection{The shape of the nonlinear power spectrum}

We complete these investigations with the non-linear evolution of
the matter density contrast power spectrum. Assuming stable clustering
ansatz and the Hamilton \emph{et al} mapping\cite{Hamiltonetal:91},
we compare the power spectrum in quintessence and cosmological constant
models in the deeply non-linear regime. 

The linear matter power spectrum in quintessence scenarios had been
studied before \cite{Fabris:PertScaQeffec:97,FerreiraJoyce:ExpQuint98,Caldwell:pipo98,RzMartinBrax:CMBQSUGRA2000}.
For modes inside the horizons, models with a pure cosmological constant
or with a quintessence field show very little differences in the shape
of the linear transfer function. Limiting our study to those modes,
we can reliably approximate the linear quintessence power spectrum
by a standard \( \Lambda  \)CDM one. 

However there is no reason for the non-linear evolution of these models
to lead to the same power spectrum. Indeed, we showed in section \ref{sec:lingrowth}
that generically the large scale structure grows more slowly in quintessence
scenarios. Hence, for the same amount of structure today, the density
contrast had to be bigger in quintessence scenario at early time.
This implies that modes that are in the non-linear regime now have
reached this regime sooner in quintessence scenarios. 

Following the idea of Hamilton et al.\cite{Hamiltonetal:91}, later
extended by Peacock and Dodds\cite{PeacockDodds:96,PeacockDodds:94},
we postulate that one can describe the effects of non-linear evolution
through a universal function \( f_{\mathrm{nl}} \) that maps the
linear power spectrum onto the non-linear one\begin{eqnarray}
\Delta ^{2}_{\textrm{nl}}(k_{\textrm{nl}}) & = & f_{\textrm{nl}}(\Delta ^{2}(k))\nonumber \\
k & = & \left( 1+\Delta _{\textrm{nl}}^{2}(k_{\textrm{nl}})\right) ^{-1/3}k_{\textrm{nl}}\\
\Delta ^{2}(k) & = & 4\pi k^{3}P(k).\nonumber 
\end{eqnarray}
 Enforcing stable clustering, the previous authors have shown that
this function must follow an asymptotic behavior at large \( x \)
such that \( f_{\textrm{nl}}(x)\propto g(\Omega )^{-3}x^{3/2} \)
--- where \( g(\Omega )=D_{1}(a)/a \), is the ratio of the linear
growth factor to the EdS growth factor described in section \ref{sec:lingrowth}
--- and have proposed analytic forms for \( f_{\textrm{nl}} \) that
depend on the cosmological parameters through \( g(a) \) only and
that are calibrated on various \( N \)-body simulations. We assume
that their results hold for quintessence scenarios. In particular,
we assume that the non-linear regime always reaches a stable clustering
regime. Moreover, and in absence of quintessence N-body simulation
for our particular scenarios, we also assume that the normalization
factor in the asymptotic branch is independent on the cosmological
scenario. These assumptions are not trivial and can probably be challenged
(see for instance \cite{Ma:QuintSpectrePuiss99} where the behavior
of the nonlinear power spectrum is investigated for effective quintessence
models with a different perspective). 

From the Hamilton et al. ansatz we expect very different behaviors
for the small scale non-linear power spectrum when quintessence and
non-quintessence models are compared. We previously obtained that
about \( 20\% \) to \( 30\% \) discrepancy is expected, depending
of the quintessence potential, between \( g^{Q} \) and \( D^{\Lambda } \).
The consequences of this a priori modest discrepancy are dramatic
for the nonlinear power spectrum. For a mode that entered the non-linear
region long before redshift \( z \), we have\begin{eqnarray}
\frac{P^{Q}_{\textrm{nl}}(k_{\textrm{nl}},z)}{P^{\Lambda }_{\textrm{nl}}(k_{\textrm{nl}},z)} & = & \frac{\Delta ^{Q}_{\textrm{nl}}(k_{\textrm{nl}},z)^{2}}{\Delta ^{\Lambda }_{\textrm{nl}}(k_{\textrm{nl}},z)^{2}}\label{eq.p3dsp3d} \\
 & \sim  & \frac{g^{Q}(z)^{-3}}{g^{\Lambda }(z)^{-3}}\, \frac{\Delta ^{Q}(k_{Q},z)^{3}}{\Delta ^{\Lambda }(k_{\Lambda },z)^{3}}\nonumber 
\end{eqnarray}
 where \( k_{Q} \) (resp. \( k_{\Lambda } \)) is the linear mode
in the quintessence (resp. \( \Lambda  \)) linear power spectrum
giving rise to the \( k_{\textrm{nl}} \) mode in the non-linear spectrum.
Assuming, as stated before, that at the sub-horizon scales we are
interested in, the linear power spectrum of the models are identical,
up to a normalization factor \( P_{0} \), we write \( P(k,z)=g^{2}(z)a^{2}\, P_{0}\, k^{n} \)
with \( n>-3 \). For a mode in the non-linear region we have\begin{equation}
k\sim \left( \frac{k_{\textrm{nl}}^{2}}{a^{2}P_{0}}\right) ^{\frac{1}{5+n}}
\end{equation}
 and Eq. (\ref{eq.p3dsp3d}) reduces to\begin{eqnarray}
\frac{P^{Q}_{\textrm{nl}}(k_{\textrm{nl}},z)}{P^{\Lambda }_{\textrm{nl}}(k_{\textrm{nl}},z)} & \sim  & \frac{g^{Q}(z)^{-3}}{g^{\Lambda }(z)^{-3}}\left( \frac{g^{Q}(z)^{2}a^{2}\, P^{Q}_{0}\, k_{Q}^{n+3}}{g^{\Lambda }(z)^{2}a^{2}\, P_{0}^{\Lambda }\, k_{\Lambda }^{n+3}}\right) ^{3/2}\nonumber \\
 & \sim  & \left( \frac{P_{0}^{Q}}{P_{0}^{\Lambda }}\right) ^{\frac{3}{2}\left( 1-\frac{n+3}{n+5}\right) }.
\end{eqnarray}
 We suppose here that the spectral index \( n \) is identical for
both \( P^{Q}(k_{Q}) \) and \( P^{\Lambda }(k_{\Lambda }) \) which
is a reasonable approximation. Finally if we set the spectrums to
both fit the cluster normalization, the ratio \( P^{Q}_{0}/P^{\Lambda }_{0} \)
is simply the ratio of the growing modes at \( z=0 \) and we get
\begin{equation}
\label{nlPkratio}
\frac{P^{Q}_{\textrm{nl}}(k_{\textrm{nl}},z)}{P^{\Lambda }_{\textrm{nl}}(k_{\textrm{nl}},z)}=\left( \frac{g^{Q}(z=0)}{g^{\Lambda }(z=0)}\right) ^{-3\, \left( 1-\frac{n+3}{n+5}\right) }.
\end{equation}
 Note that the exponent gets close to \( -3 \) when the spectral
index goes to \( -3 \). Given the variation of \( g^{Q} \) the ratio
given in Eq. (\ref{nlPkratio}) can be as large as \( 2 \)! 

The same ratio is easier to compute for modes in the linear regime,
\begin{equation}
\frac{P^{Q}(k,z)}{P^{\Lambda }(k,z)}=\left( \frac{g^{Q}(z)\, g^{\Lambda }(0)}{g^{Q}(0)\, g^{\Lambda }(z)}\right) ^{2}.
\end{equation}

These simple investigations show unambiguously and with a limited
number of assumptions that the shape of the nonlinear power spectrum
is very sensitive to the presence of a quintessence field. In order
to have a full description of the power spectrum behavior, including
the intermediate regime, we use the Peacock and Dodds prescription.
This formula has been shown to be reasonably accurate for \emph{effective}
quintessence models in N-body simulation, at least at low redshift
\cite{Ma:QuintSpectrePuiss99}. Anyway, we are more interested, in
this paper, in the general trend rather than a percent accurate description
of the transition between the linear and non-linear evolution. 
\begin{figure}[ht]
{\centering \resizebox*{0.45\textwidth}{!}{\rotatebox{270}{\includegraphics{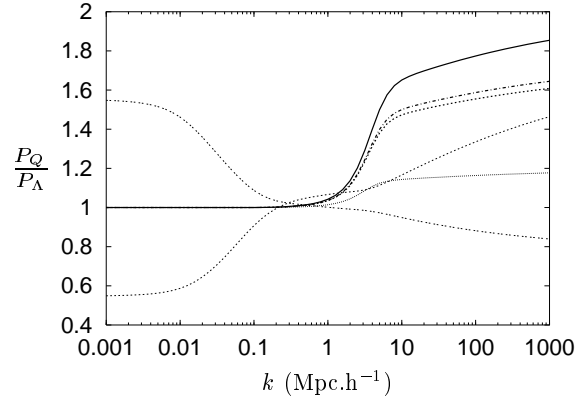}}} \par}

\caption{\label{fig.P3D}the ratio \protect\( P_{3D}^{Q}/P_{3D}^{\Lambda }\protect \)
at \protect\( z=0\protect \). The solid and dashed lines are the
SUGRA quintessence models for \protect\( \alpha =11\protect \) and
\protect\( \alpha =6\protect \), the dot-dashed one is Ratra-Peebles
and the thin dotted line is the effective quintessence model for \protect\( \omega _{Q}=-0.8\protect \).
We also show here the results for a \protect\( \Omega _{0}=0.4\protect \)
flat \protect\( \Lambda \protect \)CDM (thin dashed line) and the
\protect\( \Omega _{0}=0.25\protect \) flat \protect\( \Lambda \protect \)CDM
(thin double dashed line) models which where discussed in previous
sections.}
\end{figure}
 Fig.\ref{fig.P3D} shows a comparison between quintessence models
and a standard \( \Lambda  \)CDM. The curves represent the ratios
between a quintessence non-linear \( P_{3D} \) at \( z=0 \) and
the \( \Lambda  \)CDM one assuming the linear power spectra have
the same normalization. As expected, the small scale ratio tends toward
the \( (g_{+}^{Q}/g_{+}^{\Lambda })^{-3} \) asymptote. The shape
of the curve in this region is described by the \( 3\, \left( 1-(n+3)/(n+5)\right)  \)
power calculated above. Since the effect goes like the growth ratio
to the third, it is much smaller for \emph{effective} quintessence
where the discrepancy between its growth ratio and the \( \Lambda  \)
model growth are smaller. We stress therefore again that the use of
a constant equation of state cannot account of the amplitude of this
effect as it is expected in realistic models. 

Note that the transition between the linear and non-linear regime
depends on the ansatz used for \( f_{\textrm{nl}} \). However, a
very sharp transition, as encountered here, is not unnatural. It accounts
for different times a given mode enters the non-linear regime in different
models. If one follows a given mode throughout its evolution, it will
first obey the linear growth and evolves as \( a^{2}g^{2}(a) \).
Then it enters the nonlinear regime and grows as \( a^{3} \). The
transition between this two regimes is very sharp. When models with
a different growth factor are compared, this rapid transition translates
into a sharp increase of the power spectrum ratio between the linear
and the non-linear regime.

Moreover we note that this effect cannot be mistaken with a variation
of \( \Omega _{0} \). The latter has a much more dramatic effect
on the shape of the linear power spectrum through a change of the
shape of the transfer function. In this case, not only is the linear
growth rate changed but the position of the maximum of the linear
power spectrum is also shifted. 

In principle large-scale galaxy surveys such as the 2dF or the SDSS
should be able to put constraints on the amplitude and shape of the
power spectrum. The possible effects of biasing mechanisms, that are
extremely badly understood in the transition regime between the linear
and the non-linear regime, prevent however to have a robust and reliable
test of these scenarios. In the next section we rather try to validate
these properties in the context of a better defined observational
procedure: the projected power spectrum in weak lensing surveys.

\subsection{Projected power spectrum in weak lensing surveys}

Weak lensing surveys can potentially provide us with precision maps
of the projected density up to redshifts around one \cite{vWetal:lensSurv00,BaconRefregierEllis:LensSurv00,wittman:lensSurv00,Kaiser:lensSurv00,Maoli:lensSurv00,WKL:lensSurv01,vWetal:lensSurv01,RhodeRefDroth:lensSurv01}.
These measurements are expected to be free of observation biases once
the redshift distribution of the sources is known. 

Weak lensing surveys, through the observation of the deformation of
background galaxies, can give access to the convergence field. The
latter can be written as the projection of the matter density along
the line-of-sight\cite{BvWM:WLensTheo97}, \begin{equation}
\kappa (\vec{\alpha })=\int _{0}^{\mD _{s}}\d {\mD }w(\mD )\, \delta (\vec{\alpha },z(\mD )),
\end{equation}
 where the \( \mD  \) stands for the angular distances (and this
formula is valid for a flat spatial curvature only). The geometrical
kernel \( w(\mD ) \), defined in eq. (\ref{eq:w:def}), accounts
for the projection effects. 

We expect that the ratio of the power spectrum of the convergence
field for models with a cosmological constant and models with a quintessence
field exhibit roughly the same properties than the ratio of the three
dimensional power spectra. However, the value of this ratio is expected
to be affected by a corrective factor induced by the geometrical kernel
that itself depend on the details of the cosmological model. In the
following we present an evaluation of this rescaling factor along
the line of reasoning of the previous subsection. 

In the small angles limit, we have \begin{equation}
P_{\kappa }(\vec{\alpha })=\int _{0}^{\mD _{s}}\! \! \! \d {\mD }\, w(\mD )^{2}\int \! \! \! \frac{\d {^{2}}k}{(2\pi )^{2}}P_{3D}(k,z(\mD )){\mathrm{e}}^{\i \vec{k}\mD \cdot \vec{\alpha }}
\end{equation}
 so that the convergence power spectrum is simply \begin{equation}
P_{\kappa }(l)=\int _{0}^{\mD _{s}}\! \frac{\d {\mD }}{\mD ^{2}}\, w(\mD )^{2}P_{3D}(l/\mD ,z(\mD )),
\end{equation}
 where a possible redshift evolution of the shape of the power spectrum
is included. The kernel \( w \) is a bell shaped window that reaches
its maximum at \( z_{\textrm{eff}}=z(\mD _{s}/2) \). To evaluate
roughly the rescaling factor we will approximate \( w^{2} \) by a
simple Dirac function \( w^{2}(\mD )\sim w^{2}_{\textrm{eff}}\, \delta (\mD -\mD _{s}/2) \)
with \( w_{\textrm{eff}}=\int ^{\mD _{s}}\d {\mD }w(\mD ) \) and
\( z_{\textrm{eff}}\sim 0.4 \) (it depends on the cosmology we are
considering) for sources at redshift \( z_{s} \). Now, the ratio
of the convergence power spectrum simply writes, \begin{equation}
\label{eq:ratiopkappa}
\frac{P^{Q}_{\kappa }(l)}{P_{\kappa }^{\Lambda }(l)}\sim \left( \frac{w^{Q}_{\textrm{eff}}/\mD ^{Q}_{s}}{w^{\Lambda }_{\textrm{eff}}/\mD ^{\Lambda }_{s}}\right) ^{2}\frac{P^{Q}_{3D}(2\, l/\mD ^{Q}_{s},z^{Q}_{\textrm{eff}})}{P^{\Lambda }_{3D}(2\, l/\mD ^{\Lambda }_{s},z^{\Lambda }_{\textrm{eff}})}.
\end{equation}
 For a mode in the non-linear region, a few percent error in the position
of the mode is not significant, so that we can ignore the difference
between \( D_{Q} \) and \( D_{\Lambda } \) in the last term of the
equation. As a result \begin{eqnarray}
\frac{P^{Q}_{\kappa }(l_{\textrm{nl}})}{P_{\kappa }^{\Lambda }(l_{\textrm{nl}})} & \sim  & \left( \frac{w^{Q}_{\textrm{eff}}\, /\mD ^{Q}_{s}}{w^{\Lambda }_{\textrm{eff}}\, /\mD ^{\Lambda }_{s}}\right) ^{2}\left( \frac{a_{\textrm{eff}}^{Q}}{a_{\textrm{eff}}^{\Lambda }}\right) ^{3}\nonumber \\
 &  & \hspace {1cm}\times \left( \frac{g^{Q}(z=0)}{g^{\Lambda }(z=0)}\right) ^{-3\, \left( 1-\frac{n+3}{n+5}\right) }
\end{eqnarray}
 which, compared to Eq. (\ref{nlPkratio}), contains an extra geometrical
factor due to the projection effects. It evaluates to 0.8 to 0.9 depending
on the models and position of the source plane. We expect therefore
the conclusions reached in the previous section to survive in weak
lensing observations. 

Similarly, for a mode in the linear region, we have \begin{eqnarray}
\frac{P^{Q}_{\kappa }(l)}{P_{\kappa }^{\Lambda }(l)} & \sim  & \left( \frac{w^{Q}_{\textrm{eff}}\, /\mD ^{Q}_{s}}{w^{\Lambda }_{\textrm{eff}}\, /\mD ^{\Lambda }_{s}}\right) ^{2}\left( \frac{a_{\textrm{eff}}^{Q}}{a_{\textrm{eff}}^{\Lambda }}\right) ^{2}\nonumber \\
 &  & \, \, \, \, \, \, \, \, \, \, \, \, \times \left( \frac{g^{Q}(z_{\textrm{eff}}^{Q})}{g^{Q}(0)}\frac{g^{\Lambda }(0)}{g^{\Lambda }(z_{\textrm{eff}}^{\Lambda })}\right) ^{2}.\label{linPkratio} 
\end{eqnarray}
 Table \ref{tab.reslinp2d} gives the value of the expected ratio
in the linear region. It is about \( 0.9 \). It indicates that the
normalization ratio for the linear 3D power spectrum and the one for
the projected weak lensing spectrum will differ by this amount.

\begin{table}[ht]

\caption{\label{tab.reslinp2d}Evaluation of the ratio \protect\( P^{Q}_{\kappa }/P^{\Lambda }_{\kappa }\protect \)
in the linear domain, from Eq. (\ref{linPkratio}).}

\centering

\begin{tabular}{cccc}
&
 \( z=1 \)&
 \( z=2 \)&
 \( z=1000 \)\\
 \( \omega _{Q}=-0.8 \)&
 0.94&
 0.95&
 1.06\\
 Ratra-Peebles \( \alpha =2 \)&
 0.86&
 0.88&
 1.20\\
 SUGRA \( \alpha =6 \)&
 0.92&
 0.92&
 1.16\\
 SUGRA \( \alpha =11 \)&
 0.91&
 0.91&
 1.18\\
\end{tabular}
\end{table}

These semi-analytical results give a good account of the ratio \( P_{\kappa }^{Q}/P_{\kappa }^{\Lambda } \)
for all angular scales. At large scale, it is roughly flat; its value
is given in table \ref{tab.reslinp2d}. Then, as we get closer to
the transition between the linear and the non-linear regime, the ratio
exhibits a slight drop. Indeed, from Eq. (\ref{eq:ratiopkappa}) the
\( \Lambda  \)CDM model enters the non-linear regime before, because
of the difference between the \( z_{\textrm{eff}} \). Hence, one
expects to have, for a few modes, a \( P_{\kappa }^{\Lambda } \)
that rises quicker than its quintessence counterpart. Then, when the
quintessence power spectrum also hits the non-linear regime, the ratio
will exhibit a shape very similar to the one of the three dimensional
power spectrum ratio, with the rescaling factor computed above. 

\begin{figure}[ht]
{\centering \resizebox*{0.45\textwidth}{!}{\rotatebox{270}{\includegraphics{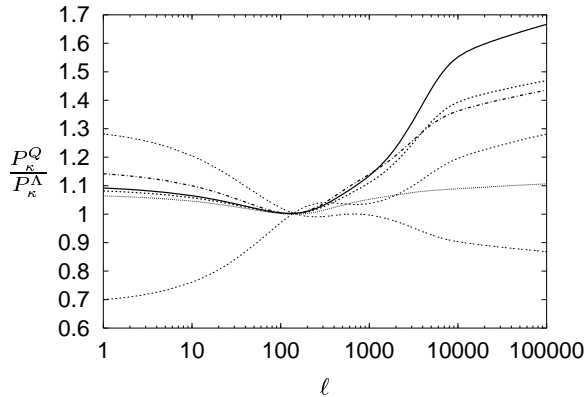}}} \par}

\caption{\label{fig.p2d}Ratios \protect\( P^{Q}_{\kappa }(l)/P^{\Lambda }_{\kappa }(l)\protect \)
for a source plane at \protect\( z=1\protect \). All models are normalized
so that \protect\( \sigma _{\kappa }\protect \) are the same at \protect\( 10'\protect \)
scale and correspond to a \protect\( \Omega _{0}=0.3\protect \),
flat \protect\( \Lambda \protect \)CDM with \protect\( \sigma _{8}=1\protect \).
The observational window, in which measurements with an accuracy better
than 10\% is foreseeable, corresponds to \protect\( \ell \protect \)
of 200 to 10,000 (minute to degree scale) where the most dramatic
changes take place. }
\end{figure}

In Fig. \ref{fig.p2d} we present the explicit computation of the
nonlinear power spectra of the convergence field using the prescription
of Peacock \& Dodds to compute the redshift evolution of the 3D power
spectrum. Unlike Fig. \ref{fig.P3D}, the power spectra are not cluster
normalized. In this case the power spectra are normalized so that
weak lensing amplitudes match at 10' scale when computed with the
linear power spectrum and match the amplitude of the recent detections
of weak lensing effects (e.g. \( \sigma _{8}\approx 1 \) for a \( \Lambda  \)-CDM
model with \( \Lambda =0.7 \)). Because of the projection effects
given in Table \ref{tab.reslinp2d}, this is not equivalent to normalized
linear 3D power spectrum. Projection effects also slightly change
the shape of the projected linear power spectrum. The redshift of
the sources is simply here assumed to be unity. The differences in
the shape of the power spectra is clearly visible and should be already
within observational constraints. 

We also give, following the same prescription for the normalization,
the effects of a change of \( \Omega _{0} \). In this case, because
we normalized to the \emph{convergence} linear power spectrum, the
effects of \( \Omega _{0} \) also directly affects the normalization.
Compared to 3D power spectra, it actually worsen the situation and
make the distinction between quintessence models and such models striking.

The result of fig. \ref{fig.p2d} gives us hope to constraint strongly
the quintessence scenario using weak lensing surveys. The next generation
weak lensing surveys will made available wide surveys where a precise
determination of the lensing effect will be possible for a range of
scale large enough to map the sharp rise predicted here. For example,
measurement of the weak lensing effect amplitude at one degree scale
and at one minute scale with only a ten percent precision appear sufficient
to test a SUGRA quintessence hypothesis. It seems that the observational
requirements are much more modest than direct measurements of the
angular distances through SNIa observations.

\subsection{Weak lensing on Cosmic Microwave Background}

In passing we note that weak lensing effects on CMB maps could be
used also to test quintessence hypothesis. Amplitude of the effects
are mainly given by the amplitude of the fluctuations of \( \kappa  \),
\( \sigma _{\kappa }^{2} \), along the line-of-sight\cite{Bernardeau:CMBLens97,SeljakZal:kappaPower99}.
In Table \ref{tab.sigmakappa} we show the amplitude of the lens effects
on the last scattering surface at two different angular resolutions.
They are mainly sensitive to the linear change of the growth rate
integrated over the line of sight. It would probably be not a crucial
test for the nature of the vacuum energy but it is potentially an
important test to pass once the general cosmological parameters will
be determined. If the coming generation of observations call for quintessence,
observation of an excess of power of the lens effect as suggested
by these calculations, would be an important consistency test.

\begin{table}[ht]

\caption{\label{tab.sigmakappa}Ratios between the amplitudes of the lens
effects on the last scattering surface for different models and the
standard \protect\( \Lambda \protect \)CDM model (\protect\( \Lambda =0.7\protect \))
at two different angular resolutions.}

\centering

\begin{tabular}{ccc}
 \( \sigma ^{2}_{\kappa } \) at \( z=1000 \)&
 5' &
 10'\\
\hline
\( \Omega =0.25 \) \( \Lambda =0.75 \) model &
 1.23&
 1.29\\
 \( \Omega =0.4 \) \( \Lambda =0.6 \) model &
 0.68&
 0.73\\
 \( \omega _{Q}=-0.8 \)&
 1.20&
 1.21\\
 Ratra-Peebles \( \alpha =2 \)&
 1.49&
 1.54\\
 SUGRA \( \alpha =6 \)&
 1.29&
 1.32\\
 SUGRA \( \alpha =11 \)&
 1.33&
 1.36\\
\end{tabular}
\end{table}

\section{Conclusion}

In this paper we have examined the growth of structure in quintessence
models in both the linear and the second order regime and present
their more striking implications for the statistical properties of
the low redshift large-scale structure of the universe. 

We paid particular attention to cases of realistic implementations
of quintessence field since they lead to scenarios where the energy
fraction in the quintessence component can represent a significant
fraction of the total energy density over a long period. We indeed
found that this effect is responsible of important differences in
the behavior of the linear growth rate of the fluctuations: For the
same values of \( \Omega _{0} \), realistic quintessence models lead
to a linear growth rate that can be 20 or 30\% lower compared to models
with a pure cosmological constant or with an effective quintessence
component (where the vacuum has a constant equation of state which
matches the angular distances constraints). 

Consequences of this discrepancy have been explored at the level of
the nonlinear power spectrum for which such differences are amplified.
For power spectra with identical linear normalization (at \( z=0 \)),
the variation of the amplitude of the nonlinear power spectrum can
be as large as 2. 

We have also computed the second order growth rate of the fluctuation.
We found that, when expressed in terms of the square of the linear
rate, it is not sensitive to the nature of the dark energy. This ratio
is actually not significantly sensitive to any of the cosmological
parameters. In this respect our result extends previously known properties. 

Weak lensing surveys appear to be the natural playground for such
effects. They combine effects on the angular distances and on the
growth rate of the fluctuations. We show that the skewness of the
convergence field, at large angular scale, is notably sensitive to
the projection effects. It is to be noted however that a universe
with quintessence field does not resemble a universe with a cosmological
constant and larger matter density (as it is the case for the behavior
of the angular distances) but rather with a lower density parameter. 

Moreover the shape of the power spectrum of the convergence field,
which identifies with a projected 3D matter power spectrum, retains
the properties found for the 3D nonlinear spectrum. It appears clearly
that CDM family models for a flat universe can be distinguished from
one another: a variation of \( \Omega _{0} \) changes the shape of
the linear power spectrum, whereas the introduction of a quintessence
field changes the time at which modes become nonlinear. 

The precision level of the current semi-analytical predictions for
the shape of the nonlinear spectrum does not permit so far to make
precise predictions from which the quintessence potential could be
reconstructed. Moreover the use of the prescription of Peacock \&
Dodds for models of quintessence with a tracking solution should probably
be validated with specific numerical simulations. 

The calculations have been done in two specific models of quintessence,
the Ratra-Peebles model and the Sugra model developed in \cite{BrxMartin:SUGRA99}.
We think however that our conclusions would survive for any model
where the energy density in the quintessence component can be a significant
fraction of the total energy up to recombination. 

\acknowledgements The authors are very thankful to J. Martin, A. Riazuelo,
Ph. Brax, L. van Waerbeke for fruitful discussions.

\end{document}